\documentclass[conference]{IEEEtran}
\IEEEoverridecommandlockouts
\usepackage{cite}
\usepackage{amsmath,amssymb,amsfonts}
\usepackage{algorithmic}
\usepackage{graphicx}
\usepackage{textcomp}
\usepackage{xcolor}
\usepackage{hyperref}
\usepackage{subfigure}

\usepackage[T1]{fontenc}
\usepackage{multirow}
\usepackage{algorithmic}
\usepackage{algorithm}
\usepackage{color}
\newcommand{\eat}[1]{}
\newcommand{\stitle}[1]{\vspace{1ex} \noindent{\emph{#1}}}

\def\BibTeX{{\rm B\kern-.05em{\sc i\kern-.025em b}\kern-.08em
    T\kern-.1667em\lower.7ex\hbox{E}\kern-.125emX}}
\begin{document}

\title{Accelerating Heterogeneous Tensor Parallelism via Flexible Workload Control}

\author{
    \IEEEauthorblockN{Zhigang Wang}
    \IEEEauthorblockA{
        \textit{Ocean University of China}\\
        wangzhigang@ouc.edu.cn}
    \and

    \IEEEauthorblockN{Xu Zhang}
    \IEEEauthorblockA{
        \textit{Ocean University of China}\\
        zhangxu@stu.ouc.edu.cn}
    \and

    \IEEEauthorblockN{Ning Wang}
    \IEEEauthorblockA{
        \textit{Ocean University of China}\\
        wangning8687@ouc.edu.cn}
    \and

    \IEEEauthorblockN{Chuanfei Xu}
    \IEEEauthorblockA{
        \textit{Huawei Technologies Co., Ltd.}\\
        xuchuanfei@huawei.com}
    \and

    \IEEEauthorblockN{Jie Nie}
    \IEEEauthorblockA{
        \textit{Ocean University of China}\\
        niejie@ouc.edu.cn}
    \and

    \IEEEauthorblockN{Zhiqiang Wei}
    \IEEEauthorblockA{
        \textit{Ocean University of China}\\
        weizhiqiang@ouc.edu.cn}
    \and

    \IEEEauthorblockN{Yu Gu}
    \IEEEauthorblockA{
        \textit{Northeastern University}\\
        guyu@mail.neu.edu.cn}
    \and

    \IEEEauthorblockN{Ge Yu}
    \IEEEauthorblockA{
        \textit{Northeastern University}\\
        yuge@mail.neu.edu.cn}
}

\maketitle

\begin{abstract}
Transformer-based models are becoming deeper and larger recently. For better scalability, an underlying training solution in industry is to split billions of parameters (tensors) into many tasks and then run them across homogeneous accelerators (e.g., GPUs). However, such dedicated compute cluster is prohibitively expensive in academia and moderate companies. An economic replacement is to aggregate existing heterogeneous devices and share resources among multi-tenants. Nevertheless, static hardware configurations and dynamic resource contention definitely cause straggling tasks, which heavily slows down the overall training efficiency. Existing works feature contributions mainly tailored for traditional data parallelism. They cannot work well for the new tensor parallelism due to strict communication and correctness constraints.

In this paper we first present ZERO-resizing, a novel dynamic workload balancing technique without any data migration. We tune workloads in real-time by temporarily resizing matrices involved in core tensor-related computations. We particularly design data imputation and priority selection policies to respectively satisfy consistency constraint required by normal training and reduce the accuracy loss. We also give a lightweight data migration technique without loss of accuracy, to cope with heavy heterogeneity. Our final SEMI-migration solution is built on top of these two techniques and can adaptively distinguish their respective balancing missions, to achieve an overall success in efficiency and accuracy. Extensive experiments on the representative Colossal-AI platform validate the effectiveness of our proposals.
\end{abstract}

\begin{IEEEkeywords}
Tensor Parallelism, Heterogeneous Environments, Dynamic Workload Balancing, Data Migration
\end{IEEEkeywords}

\section{Introduction}\label{section:1intro}
Since the public release of ChatGPT, hundreds of foundation models have been designed. They are experiencing rapid growth and widespread adoption. Among them, most are transformed-based~\cite{transformer}, like Llama2 for natural language processing~\cite{llama2}, Vision Transformers (ViT) for image understanding~\cite{vit}, and DALL$\cdot$E2 for creating images from texts~\cite{dalle2}.

Training foundation models with billions of parameters (tensors) is significantly memory-consuming, and hence multi-accelerators (e.g., GPUs) are required to provide aggregated resources. However, different from regular models where intermediate data dominate the memory footprint (like activations in CNN), now parameters as well as associated gradients and optimizer states also generate considerably huge space requirements. The early data parallelism solution~\cite{dp} cannot work here because it fully replicates parameter-related variables on each device and then the memory capacity can be easily exceeded. To crack this nut, a prominent alternative is to directly split tensors across several tasks, so that the device running a task can accommodate assigned data well, termed as {\bf Tensor Parallelism}~\cite{r4}.

A transformer architecture typically consists of tens or even hundreds of stacked multi-head attention and feedforward networks. Under tensor parallelism, frequent global data collection is performed by synchronizing all tasks, to provide correct inputs for the two components. Besides, their respective linear projections and transformations involve compute-intensive matrix multiplications. The frequent synchronization operations and heavy computation workloads make the parallel training efficiency very sensitive to each task's speed.

Top IT companies have made huge investments in building AI compute clusters equipped with homogeneous accelerators, like AI Research SuperCluster (RSC) in Meta. Then all tasks can run at the same speed to eliminate any possible waiting cost. However, such prohibitively expensive cluster is out of reach for most AI scientists in academia and moderate companies. An economic replacement is to aggregate existing heterogeneous devices. Worse, even for the dedicated RSC, the number of daily submitted training jobs is up to 35,000. The static hardware configurations and dynamic multi-tenant resource contention on the congested cluster together cause heavy heterogeneity. Thus, parallel tasks usually proceed at very different speeds. As a result, the aggregated compute power cannot translate to high-speedup numbers in practice.

We are aware that many researchers have studied such straggling problem. Their techniques make significant advancements but most of them are for traditional data parallelism, which cannot be directly extended to tensor parallelism. They include static skewed data assignment~\cite{r24,r25,r26}, which still suffers from dynamic and temporary stragglers; data migration~\cite{r16,r17,r18}, which incurs additional runtime latency; relaxed synchronous constraint with stale intermediate results~\cite{r7,r9,r13,r31}, where frequent data collection yields cascading error accumulation and finally impairs the model accuracy. Recently some works attempt to dynamically tune workloads of tasks, by customizing the numbers of their original inputs in each iteration (i.e., batch size)~\cite{r21,fsp}. That generates skewed contributions for parameter refinement since data assigned onto tasks are scanned with different frequencies. Also, under tensor parallelism, this design accordingly changes the size of split tensor and then yields communication failures.

Clearly, there is an imperative need for efficient tensor parallelism in dynamic heterogeneous environments. Inspired by customizing batch size and data migration, this paper pursues such target in a more sound and effective manner.

We firstly propose {\bf ZERO-resizing}, which temporarily resizes matrices involved in tensor computations, so as to dynamically tune workloads on demand. Specifically, once the straggling phenomenon happens, we immediately prune some columns or rows (dual) in matrices to reduce workloads. Then stragglers can roughly proceed as fast as normal tasks. Pruning is performed in both forward- and backward-propagations. But in the latter, the dimension of output matrices will change, terminating propagation abnormally. We thereby recover the dimension by imputing missing data, to enforce the consistency constraint on dimensions. That makes our pruning different from existing network pruning in inference where parameters are completely and permanently removed~\cite{r27,r28}. We particularly employ a lineage lookup table to record pruned and recovered dimension locations. It helps to exactly match gradients with weights and then correctly refine the latter. Other key issues are also studied, including how many columns/rows and which column/row should be pruned. Our solution can adaptively react to the change of heterogeneity, and then prioritize data selection based on importance evaluation and the residing layer.

ZERO-resizing avoids migration costs, but pruning inevitably impairs the model accuracy. We then propose to migrate workloads from stragglers to fast tasks, and tackle the runtime latency head on. We select high-throughput broadcasting communication primitives \emph{broadcast-reduce}, instead of conventionally used \emph{scatter-gather}, by in-depth analysis and extensive tests. We particularly analyze the \emph{sending-collecting} migration dataflow. We transform some existing global \emph{reduce} operations into local variants, and then merge the latter into normal data collections, to eliminate redundant communication behaviors. That further reduces additional migration costs. However, when many tasks become stragglers, this {\bf lightweight} design is still cost-ineffective since benefits brought by enhancements mentioned above decrease.

In general, resizing and migration have different favorite scenarios, but neither of them always works best in all cases. That motivates us to design a hybrid solution called {\bf SEMI-migration}, on top of them, to strike a good balance between accuracy loss and runtime latency. It solely runs resizing if the speed difference is tolerable; and otherwise re-assigns workloads via migration, to make sure that resizing can help stragglers with emigrated data to easily catch up with normal tasks with newly immigrated data. More generally, when encountering many stragglers, it first sorts them in descending order of their speed differences. Afterwards, the top-stragglers employ migration, while others employ resizing. For the last too heavy, and multi-straggling scenarios, we formulate the balancing mission allocation question. We attempt to answer it by real-time statistics, to roughly hit the ``sweet spot'' between accuracy and efficiency.

The major contributions are summarized below.
\begin{itemize}
    \item Proposing a novel matrix-resizing approach \emph{ZERO}-resizing, to dynamically balance workloads under tensor parallelism, which employs data imputation and priority selection policies to guarantee consistency in computations and mitigate the loss of accuracy.
    \item Proposing another lightweight migration approach to cope with heavily heterogeneous environments, which reduces the migration runtime latency via existing tree-based broadcasting/reducing efforts and our novel reducing merging optimization.
    \item Building a hybrid solution \emph{SEMI}-migration on top of the two approaches, which can smartly run the reasonable one in different scenarios based on the cost-benefit analysis.
    \item Performing extensive experimental studies using a well-known foundation model ViT with billions of parameters. We integrate our proposals into the up-to-date training platform Colossal-AI~\cite{r5}. Compared against the original version, our enhanced variant improves the efficiency by 18.5\% and 77.6\% with negligible accuracy loss, respectively in homogeneous and heterogeneous environments.
\end{itemize}

Below, Sec.~\ref{section:2pre} briefly overviews tensor parallelism and analyzes the challenges in heterogeneous environments. Sec.~\ref{section:3zero} and Sec.~\ref{section:4semi} present the details of our balancing solutions. Sec.~\ref{section:5exp} gives experiment results. Sec.~\ref{section:6related} summarizes related works. Finally, Sec.~\ref{section:7con} concludes this paper.

\section{Preliminaries}\label{section:2pre}
\begin{figure*}
    \centering
    \includegraphics[width=0.90\linewidth]{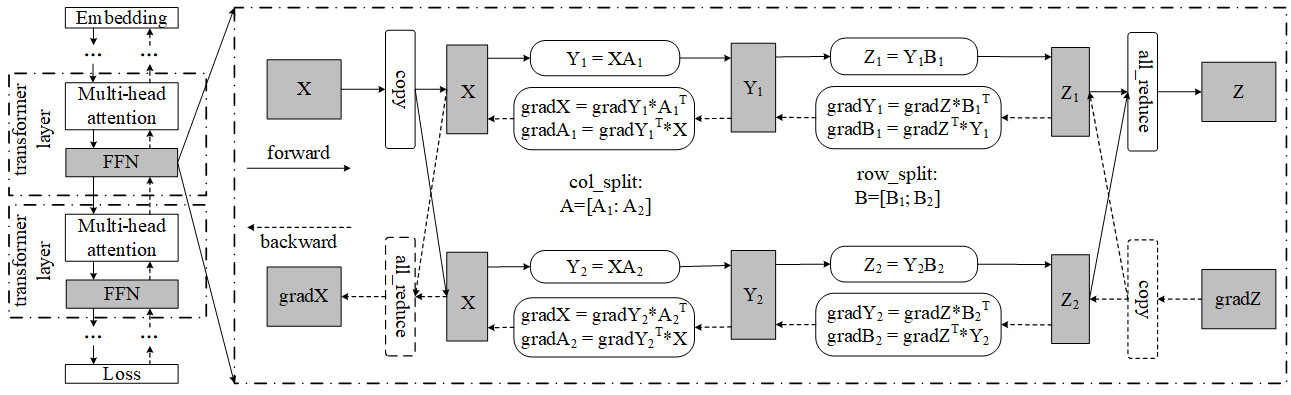}
    \caption{Tensor parallelism in a FFN layer}
    \label{figure:1}
\end{figure*}

Under model parallelism, each task/device has only a portion of parameters. That reduces the storage burden and hence is particularly suitable for training foundation models with billions of parameters. Currently, it is divided into tensor parallelism and pipeline parallelism. Pipeline parallelism involves inter-layer partitioning, where the parameters (tensors) are split among tasks in a layer-centric manner. In contrast, tensor parallelism employs intra-layer partitioning. It distributes parameters within a layer onto different tasks. In comparison to inter-layer partitioning, intra-layer partitioning avoids bubbles (the waiting cost between tasks in the pipeline queue). Thus, it can fully utilize compute power, but at expense of frequently synchronizing tasks for collecting tensor computation results. Frequent synchronization is sensitive to heterogeneous environments, which is the focus of this paper.

\subsection{Tensor Parallelism in Transformer}\label{section:2pre:tensor}
The transformer model mainly consists of stacked multi-head attention layers and feedforward networks (FFN). Its core operations are matrix multiplications between input and weight matrices. Below we introduce the specific implementations along with its two modes, namely, row- and column-wise splitting. Fig.~\ref{figure:1} illustrates the training process using the FFN layer as an example. A FFN layer consists of two linear layers. Given an input matrix $X$ and weight matrices $A$ and $B$, during the forward propagation, intermediate outputs $Y$ and final output $Z$ are computed. Afterwards, in the backward propagation, gradients for $X$, $A$, and $B$ are calculated, based on preserved intermediate results and gradients from the previous layer. Take $Y\!=\!XW$ as an example. Assume the tensor parallelism is 2. In the column-wise mode, $X$ is duplicated with two identical copies, $X_1$ and $X_2$; while $W$ is physically split into $W_1$ and $W_2$ by columns. Therefore, we require an all-gather communication to obtain the forward propagation result $Y\!=\![X_1W_1, X_2W_2]$, and an all-reduce communication for the gradient $dL/dX\!=\!dL/dX_1\!+\!dL/dX_2$ in backward propagation. In the row-wise mode, $X$ is split by columns into $X_1$ and $X_2$, and $W$ is partitioned by rows into $W_1$ and $W_2$. Then we run all-reduce for $Y\!=\!X_1W_1\!+\!X_2W_2$ in forward propagation, and all-gather for $dL/dX\!=\![dL/dX_1, dL/dX_2]$ in backward propagation.

\subsection{Communication- and Computation-intensive Behaviors}\label{section:2pre:complexity}
We clearly see that tensor parallelism has complex partitioning and communication patterns. No matter row- or column-wise splitting mode, two communication operations are required to collect partial results of parallel matrix multiplication. Currently, the widely used classic 1D tensor parallelism column splitting is used in the first linear layer, and row splitting is used in the second. Then each FFN layer only requires one all-reduce operation for collection in forward or backward propagation. On the other hand, matrix multiplication is compute-intensive, and it is run several times in the multi-head attention and FFN layers. Worse, deep learning typically requires refining model parameters again and again by scanning the total input data with hundreds of epochs. For quickly propagating updated parameters, we batch the processing of input samples by evenly partitioning them into fine-grained iterations. An epoch thereby consists of multiple iterations and each of the latter will run stacked transformer layers in both forward- and backward-propagations. As a result, the compute-intensive multiplication and communication-intensive all-reduce operations are definitely run with significantly high frequency.

For better understanding the complexities, we next give detailed introduction. Because linear layers only differ in hidden size, without loss of generality, this paper uses one type of linear layer in the column-wise tensor parallelism as an example for illustration. Assume that the tensor parallelism is $e$ (the number of parallel tasks), the batch size is $bs$, the length of an input sample sequence is $sql$, and the size of hidden layers is $hs$. In forward propagation, the dimensions of the \emph{input} matrix are $[bs, sql, hs]$, and $[hs/e, hs]$ for the \emph{weight} matrix. For the \emph{output} matrix based on \emph{input} and transposed \emph{weight}, they are $[bs, sql, hs/e]$. In backward propagation, taking \emph{grad\_weight} as an example, it requires \emph{input} and \emph{grad\_input}. The dimensions of the latter is $[bs, sql, hs/e]$. \emph{grad\_input} is also transposed and then multiplied by \emph{input}. The resulting \emph{grad\_weight} thereby has $[hs/e, hs]$ dimensions. Similarly, in backward propagation, another important dataflow is for \emph{grad\_input}/\emph{grad\_output}. It is obtained by multiplying its old version from the previous layer by the transposed \emph{weight}.

Take ViT with $hs\!=\!2048$ and $depth\!=\!24$ (the number of stacked transformer layer) as an example. Its total number of parameters is up to 1.2 billion. If we train it under 1D tensor parallelism, each iteration requires 48 all-reduce operations for the multi-head attention and FFN layers, to synchronize tasks for result collection. Assume that the number of epochs is 150 and each further consists of 10,000 iterations. The entire training process requires up to 144 million all-reduce/synchronization operations. In terms of the matrix computation complexity, we use an example \emph{output} in forward propagation of the multi-head attention layer. Assume that $bs\!=\!64$ and $sql\!=\!65$. The complexity in one iteration $\Omega(bs\!\times\!sql\!\times\!hs\!\times\!hs)$, easily exceeds tens of billions of floating-point operations.

\subsection{Challenges in Heterogeneous Environments}\label{section:2pre:challenges}
Heterogeneous compute environments are usually encountered in real-world. On the one hand, limited by the economic budget, devices are purchased at very different times. Note that hardware architectures are frequently updated; while old devices exhibit performance degradation due to natural usage loss. However, the huge compute and memory requirements of foundation models enforce us to run training on these heterogeneous accelerators~\cite{r23,r24,r25}. Their compute power differences can be significantly large. For example, we achieve up to 19.5 TFLOPS power for NVIDIA A100, but only 9.3 TFLOPS for P100. Then the compute-intensive matrix multiplication will proceed at very different speeds, on the two accelerators. Together with highly frequent synchronization operations, it is difficult to translate aggregated power into the overall training speedup. On the other hand, no matter private or public clouds equipped with many devices, there exist multiple tenants to share hardware resources. Resource contention incurs heavily dynamical heterogeneous feature. General-purpose optimizations, represented by container-based resource isolation, cannot work well, especially for compute-intensive applications typically like matrix multiplication~\cite{cgrouptest}. Worse, considering the overall resource utilization and user experience, these techniques are usually executed in a compromised manner in practical tools (e.g., Yarn).

In summary, accelerating tensor parallelism in heterogeneous environments is urgently needed. This paper focuses on this issue and gives our solutions in Sec.~\ref{section:3zero} and Sec.~\ref{section:4semi}.

\section{ZERO-resizing: Balancing Workloads by Resizing Matrices}\label{section:3zero}
This section describes ZERO-resizing, a scheme designed for workload balancing by pruning matrix computations in linear projection and transformation layers in a priority manner. Below, we first give the dual pruning policy tailored to tensor parallelism, and then present our priority pruning principles.

\subsection{Dual-pruning and Imputation}\label{section:3zero:pruning}

\begin{figure*}
    \centering
    \includegraphics[width=0.8\linewidth]{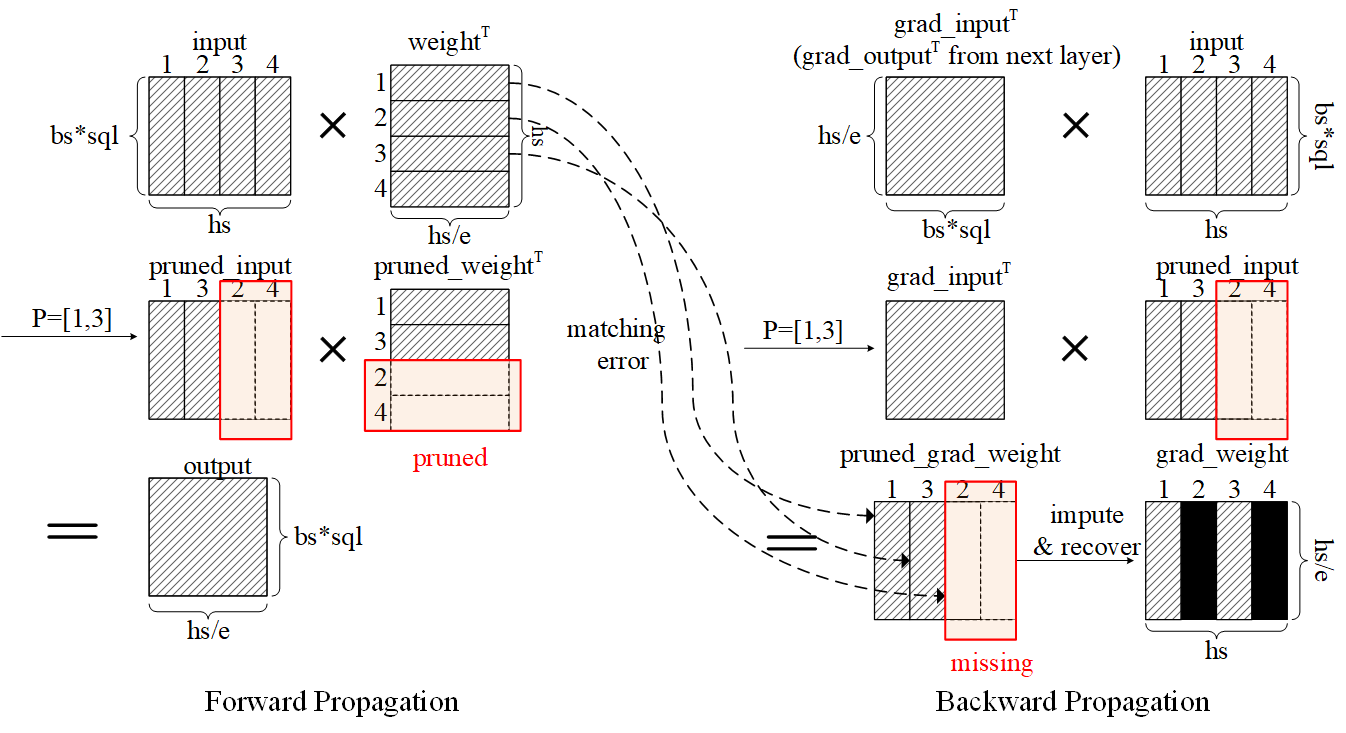}
    \caption{Matrix pruning and imputation process ($\gamma=0.5)$}
    \label{figure:2}
\end{figure*}

In heterogeneous environments, the impact of straggling tasks on the overall efficiency is mainly dominated by matrix multiplication workloads. Clearly, if we can reduce the workloads on stragglers on demand, then all tasks can be synchronized without any waiting costs. Since the workloads are associated with the dimensions of input matrices, we can achieve this goal by dynamically resizing them. Below we overview our design, discuss some key issues for guaranteeing normal training, and finally give a preferred resizing setting to react to dynamic heterogeneity.

\stitle{Overview Resizing Design.}
In particular, we design a dual-pruning scheme for dimension reduction, based on the row-splitting and column-splitting patterns used in tensor parallelism. More specifically, during both forward and backward propagations, we perform the same pruning operations on submatrices generated by the two splitting patterns. We denote by $\gamma$ the pruning ratio of pruned dimensions to original ones. Then the remaining ratio is ($1\!-\!\gamma$).

For forward propagation, before specific multiplication, we prune ($hs\!\cdot\!\gamma$) columns from the \emph{input} matrix and \emph{weight} matrix, if the corresponding task is a straggler. Afterwards, the remaining columns are concatenated in lexicographical order. The resulting dimensions thereby become $[bs, sql, hs\!\cdot\!(1\!-\!\gamma)]$ for \emph{pruned\_input}; and $[hs/e, hs\!\cdot\!(1\!-\!\gamma)]$ for \emph{pruned\_weight}. Under this design, the dimensions of \emph{output} matrix are still $[bs, sql, hs/e]$, the same as the unpruned version, which guarantees the normal computations in subsequent layers. While, the multiplication workloads has been reduced by $\gamma$.

Backward propagation can be complicated. Now there exist two outputs of matrix multiplications, namely, \emph{grad\_output} and \emph{grad\_weight}. We use the latter as an example to show how to perform pruning operations, since they have the same dataflow operations. Here, for the \emph{input} matrix, we also pre-prune dimensions and then concatenate remained ones, based on $\gamma$, resembling pre-processing in forward propagation. But differently, for \emph{grad\_input}, i.e., \emph{grad\_output} from the previous linear layer, neither rows nor columns can be pruned. Once some rows are pruned, the multiplication of its transposed result \emph{pruned\_grad\_input}$^T$ with $[bs\!\cdot\!(1\!-\!\gamma), sql, hs/e]$ and \emph{pruned\_input} with $[bs, sql, hs\!\cdot\!(1\!-\!\gamma)]$ will definitely fail. Worse, that also prevents partial contents in the same batch from gradient calculations, in the specific layer, which possibly yields biased parameter refinement. On the other hand, if columns are pruned, the similar failure happens because \emph{grad\_input} is also involved in another dataflow, i.e., computing \emph{grad\_output} for the next linear layer. As a result, we multiply the normal-sized \emph{grad\_input} by pruned \emph{input}, and then obtain the output \emph{grad\_weight} with $[hs/e, hs\!\cdot\!(1\!-\!\gamma)]$. Similarly, in another dataflow, the dimensions of \emph{grad\_output} are also changed to $[bs, sql, hs\!\cdot\!(1\!-\!\gamma)]$.

Fig~\ref{figure:2} gives concrete examples to demonstrate how to prune dimensions in forward and backward propagations. Assume that $hs\!=\!4$ and $\gamma\!=\!0.5$. In the left subfigure (forward), \emph{input} and \emph{weight} are pruned by removing columns 2 and 4. This resulting pruned versions then have fewer dimensions but their \emph{output} still has normal dimensions. In this way, only half dimensions are involved in multiplications, and hence the workloads decrease by 50\%. In the right subfigure (backward), the commonly used \emph{input} also prunes columns 2 and 4, and then is multiplied by unpruned \emph{grad\_input}. Columns 2 and 4 are absent from the final output \emph{grad\_weight}. The workload reduction is also 50\%.

\stitle{Consistency Constraint.}
To guarantee continuous training, we enforce consistency constraint on resizing. That is, both input and output of matrix multiplication must have the same dimension size with those in normally unpruned computations. The pruned inputs are temporarily used in multiplication, and the output is recovered to the normal size if columns or rows are missing. The reasons are twofold. First, the straggling phenomenon usually temporarily happens in real environments. Permanent resizing in some layers of course can reduce workloads and then avoid waiting costs. But after a while the straggling phenomenon might disappear; the aggressively pruned dimensions cannot be recovered since related elements have been completely discarded. Second, there exists complex dependency between layers and forward-backward propagations. Missing some dimensions leads to computation failures. As shown in Fig~\ref{figure:2}, \emph{output} in forward propagation is naturally kept unchanged. Thus, such failures only happen in backward propagation, which can be further divided into two cases. One is the weight refinement where \emph{grad\_weight} with fewer columns cannot update \emph{weight} with normal-sized dimensions. The other is that multiplication requires normal-sized \emph{grad\_input}, as analyzed above, but the de facto output from the previous linear layer is not. Below, we focus on solving such dimension missing problem.

Our solution is to add missing dimensions. We still use \emph{grad\_weight} calculation as the example. We add the missing ($hs\!\cdot\!\gamma$) columns into the output \emph{grad\_weight} with $[hs/e, hs\!\cdot\!(1\!-\!\gamma)]$. And for the new \emph{grad\_output} used in the next layer, the same operation is also performed. However, now a key issue is how to correctly recover the dimension size. As shown in the right subfigure of Fig~\ref{figure:2}, the output \emph{pruned\_grad\_weight} is a matrix with $hs\!\cdot\!(1\!-\!\gamma)$ columns. However, we do not know which specific column is missing, unless we have prior-knowledge about what columns have been pruned in \emph{input}. Besides, the latter is also not directly available. This is because although the matrix is commonly used in forward propagation and backward propagation, the usage time instances are very different. We can preserve the pruned version in forward propagation and then directly use it in backward propagation. But the limited memory resource is heavily wasted. Our solution is to maintain a lineage with a series of three tuples
<\emph{layer\_name}, \emph{matrix\_name}, $P$>, where $P$ is a set of indexes related to pruned columns. Accordingly, as shown in the right bottom in Fig~\ref{figure:2}, we can re-organize columns to recover \emph{grad\_weight}. And further, the matching error during weight refinement can be avoided, since now we can correctly map the $i$-th column gradients to the $i$-th column weight parameters.

\begin{figure}[htbp]
	\centering
    \renewcommand{\thesubfigure}{\scriptsize (\alph{subfigure})\space}
	\subfigure{
			 \includegraphics[width=0.5\linewidth]{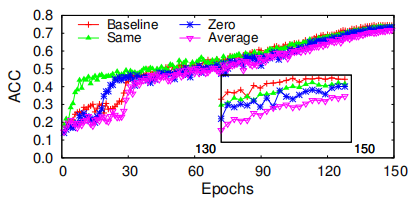}
	}
	\centering
	\caption{Impact of different imputation policies on the model accuracy (\emph{ACC})}
	\vspace{-0.2cm}
	\label{figure:imputation}
\end{figure}

The accompanied key issue is what value should be imputed into missing dimensions. Many different imputation policies can be used. We can use the same values calculated in previous iteration (\emph{Same}), the average from unpruned dimensions in the current iteration (\emph{Average}), and the uniform zero value (\emph{Zero}). We test the three policies by running the ViT-1B foundation model on Colossal-AI\footnote{The detailed settings are given in Fig.~\ref{figure:hetefixed1b} in Sec.~\ref{section:5exp:zero}.}. In addition, we set $\gamma=0.5$. In Fig.~\ref{figure:imputation}, We clearly see that \emph{Same} has the best accuracy but at expense of expensive storage costs. \emph{Zero} beats \emph{Average}. We guess the reason is that such a setting minimizes the impact of missing dimensions. Our final choice is thereby the compromised \emph{Zero} which balances space complexity and accuracy.

\stitle{Determining Pruning Ratio.}
Recall that a large pruning ratio $\gamma$ can significantly reduce workloads, but the negative impact is that many missing dimensions are imputed by zero, leading to a big accuracy loss. Thus, $\gamma$ becomes critical to balance the heterogeneous-tolerance ability and the accuracy loss. Here, our principle is to guarantee our primarily concerned goal of eliminating waiting costs, with $\gamma$ as small as possible. Before introducing the detailed design, we first state that because of the complex dependency between forward- and backward- propagations, the minimum implementation granularity is iteration. In another word, the pruning setting changes in iteration-level. Our main idea is to compare the locally recorded runtime $T_i^j$ of task-$i$ with the average of all $e$ tasks $T_{avg}$, at the $j$-th iteration. If $T_i^j\!>\!T_{avg}$, the $i$-th task judges itself as a straggler, and then uniformly calculates $\gamma_i^j$ for every transformer layer, so that the aggregated workload savings can offset the runtime gap ($T_i^j\!-\!T_{avg}$). Eq.~(\ref{equation:1}) mathematically tells us how to calculate $\gamma_i^j$, where $M_{i}^{j}$ is the runtime cost of all matrix multiplications within the $j$-th iteration.

\begin{equation}
\gamma_i^j = \frac{1}{M_{i}^{j}}\cdot(T_i^j - \frac{1}{e}\sum_{i=1}^{e}T_i^j)
\label{equation:1}
\end{equation}

Among all real-time statistics involved in Eq.~(\ref{equation:1}), $T_i^j$ and $M_{i}^{j}$ can be easily obtained at the end of an iteration. However, $T_{avg}$ is time-consuming because of the all-reduce communication. Here, instead of frequently updating $T_{avg}$, we let each task actively monitor the change of its own runtimes across already completed iterations. Once a significant change typically like an over-10\% increase, we passively update $T_{avg}$ on demand.

\subsection{Priority Pruning Principles}\label{section:3zero:priority}
Fig.~\ref{figure:imputation} reveals that resizing inevitably causes accuracy loss. Our basic resizing solution has made efforts by avoiding an over-large $\gamma$ setting and using zero to impute missing dimensions. However, zero-imputation is not loss-free as reported. Also, although the \emph{output} matrix in forward propagation has the normal dimension size, its specific element misses partial products from pruned columns. Thus, there is still huge space to further improve accuracy.

Since all possible losses are caused by pruning operations, what should be pruned becomes a critical issue. However, Sec.~\ref{section:3zero:pruning} only tells us how many dimensions should be pruned. It enforces the system to blindly and randomly select pruning data. That cannot control the impact on the direction and magnitude of gradient descent during training, especially for a large ratio $\gamma$ in heavily heterogeneous environments. In the following, we outline our priority technique and emphasize two important principles in detailed implementations.

\stitle{Priority Pruning.}
We observe that weight parameters with small gradient variations typically have a relatively marginal impact on the subsequent training rounds. Moreover, the trend of gradient changes is roughly identical to that of weight parameter changes. Inspired by those, we dynamically prioritize weight dimensions so that the one with small variation can be pruned in high probability. Given the weight matrix with $L$ columns, we then establish a priority list \emph{pri\_list} to maintain indexes of yet-to-be-pruned dimensions, with size $L_{pri}\!=\!L\!\cdot\!\gamma$. Also, a \emph{w\_var\_list} is required to keep track of weight changes where the $i$-th element $\delta_i^j$ indicates the average change of the $i$-th column, after the $j$-th update of statistics. Note that another \emph{input} matrix shares the \emph{pri\_list}. This because if \emph{weight} prunes the $i$-th column, \emph{input} should also discard the corresponding relationship by removing its $i$-th column. Finally, resembling Eq.~(\ref{equation:1}), frequently updating statistics generates substantial costs. We thereby update them in a coarse-grained epoch granularity.

\stitle{Incremental Priority Update.}
In theory, we should periodically and completely update \emph{w\_var\_list} and hence \emph{pri\_list}. However, our in-depth analysis reveals that such implementation make the priority selection fall into an endless loop. Assume that at the $t$-th epoch, we prune the $i$-th column of the weight matrix, due to its small variation. As shown in the right subfigure in Fig.~\ref{figure:2}, that means the corresponding column disappears in the pruned gradient matrix \emph{pruned\_grad\_weight}. The imputed zero values in recovered \emph{grad\_weight} will generate marginal- and even zero-impact on the $i$-th weight column, according to different optimizers. That in turn yields a small variation, and this column will be pruned in the next epoch in very high probability. Such false-positive phenomenon caused by zero-imputation enforces partial columns to be frequently pruned, once they are pruned at some epoch. It is equivalent to permanent modification of the model structure and hence affects the accuracy. We terminate this endless loop by incrementally updating statistics. In particular, at the end of each epoch, elements in \emph{w\_var\_list} related to pruned columns will be preserved. Others are normally updated by comparing the newly refined values against their old versions. Also, as refinement proceeds, weight elements generally converge to the local optimum. Thus, the variation of weights really refined will definitely become small and then the corresponding dimensions can also be selected for pruning. In theory, that builds a round-robin yet prioritized scheduling mechanism.

\stitle{Differentiated Pruning Ratios.}
Recall that in Eq.~(\ref{equation:1}), we give a uniform pruning ratio for all layers. However, the numerical values of \emph{weight} and \emph{input} matrices can be very different across layers. That motivates us to differentiate pruning ratios $\gamma_{k}^{j}$ specific to the $k$-th layer. Now $\gamma_{k}^{j}$ is solely determined by the layer-based weight variation $\delta_{k}^{j}$. More specifically, the $i$-th column residing on layer-$k$ is added into the pruning candidate set if $\delta_{ik}^{j}\!<\!\theta\!=\!N_{iter}\!\cdot\!\theta_{iter}$. Here the threshold $\theta$ is computed by the number of iterations $N_{iter}$ within an epoch and a micro-threshold $\theta_{iter}$. The latter is set as $10^{-3}$ by default. By the candidate set size, we can immediately infer $\gamma_{k}^{j}$. But the de facto pruning ratio should be equal to $\max\{\gamma_{k}^{j},\alpha\!\cdot\!\gamma^{j}\}$, to guarantee that we can satisfy the requirement of tolerating heterogeneity. Here the decay factor $\alpha$ is set as 0.8 by default to partially adjust pruning budgets among layers. That is, layers with large variations can be pruned fewer dimensions; while others with small variations will be pruned more.

Algorithm~\ref{algorithm:zero} introduces the process of ZERO-resizing for straggler in the cluster. Here, $R$ represents the row length of the parameter matrix, and $L$ represents the column length of the parameter matrix. During the training process of the $t$-th epoch, the first step involves calculating the minimum pruning ratio $\gamma^{t}$ adapted to the heterogeneity degree using Eq.~(\ref{equation:1}) (lines 1-2). Subsequently, we employ a differential pruning scheme to compute different pruning ratios $\gamma_{k}^t$ for each layer of the transformer, constructing a priority list $pri\_list_{k}^t$ based on these ratios (lines 3-15). During forward and backward propagations, temporary resizing and imputation of matrices are carried out using $pri\_list_{k}^t$ to reduce computation workloads (lines 16-22).

\begin{algorithm}
\caption{ZERO-resizing Execution($t$-th epoch)}
\label{algorithm:zero}
\textbf{Input:} $\alpha$, $\theta$, $T^t$, $e$, $N_{layer}$, $N_{iter}$, $L$, $R$,  \\
$weight^t=[w^t_1,w^t_2,...,w^t_L]$
\begin{algorithmic}[1]
\STATE $T_{avg}^t$ = {all-reduce}$(T^t)/e$
\STATE Compute $\gamma^{t}$ by Eq.~(\ref{equation:1})
\FOR{$k=1$ to $N_{layer}$}
    \STATE $\delta_{k}^t=\cup_{i=1}^{L_k}\sum_{j=1}^{R_k}|w_{ji}^t-w_{ji}^{t-1} |/R_k$
    \STATE sort $\delta_{ik}^t$ in descending order as $w\_var\_list^t$
    \FOR{$index$ not in $pri\_list_{k}^{t-1}$}
        \STATE $w\_var\_list_{index}^t= w\_var\_list_{index}^{t-1}$
    \ENDFOR
    \STATE $L_{uni}=\sum_{i=1}^{L_k}\delta_{ik}^t>\theta$
    \STATE $\gamma_{k}^t=(1-L_{uni}/{L_k})$
    \STATE $\gamma_{k}^t=\max\{\gamma_{k}^{t},\alpha\!\cdot\!\gamma^{t}\}$
    \STATE $L_{pri}=L_k*(1-\gamma_{k}^t)$
    \STATE $pri\_list_{k}^t$ = top-$L_{pri}$SelectIndex($w\_var\_list$)
    \STATE ascendSort($pri\_list_{k}^t$)
\ENDFOR
\FOR{$j=1$ to $N_{iter}$}
    \FOR{$k=1$ to $N_{layer}$}
        \STATE resize matrix during FP
        \STATE resize and impute matrix during BP
        \STATE update $weight$
    \ENDFOR
\ENDFOR
\end{algorithmic}
\end{algorithm}

\section{SEMI-migration: A Hybrid Balancing Solution}\label{section:4semi}
This section presents a hybrid load balancing solution SEMI-migration on top of already designed ZERO-resizing and a new lightweight workload migration scheme.

\subsection{Lightweight Workload Migration}\label{section:4semi:migration}
Matrix pruning can reduce the multiplication workloads on straggling tasks, but it comes with the drawback of accuracy loss. Even with the enhancement of priority pruning, it is still not loss-free. Dynamically re-balancing workloads is a classic solution without any accuracy penalty. However, the expensive data migration cost becomes an obstacle. This subsection thereby focuses on lightweight optimizations.

Once a straggler is detected as mentioned in Sec.~\ref{section:3zero:pruning}, we can use the similar method as determining $\gamma$ to calculate how many columns should be migrated. Such data will be evenly distributed across other normal tasks. A complete migration process includes sending inputs from the straggler (sender) to the target (receiver), performing computations on the latter, and finally collecting results. Recall that for each linear layer, there totally exist three matrix multiplications, respectively for \emph{output}, \emph{grap\_output}, and \emph{grad\_weight}. We thereby need to run the sending-collecting migration three times in forward- and backward-propagations. That heavily burdens tasks, especially for the straggling sender. Selecting reasonable communication primitives becomes critical to reducing migration costs.

\begin{figure}
    \centering
    \includegraphics[width=0.94\linewidth]{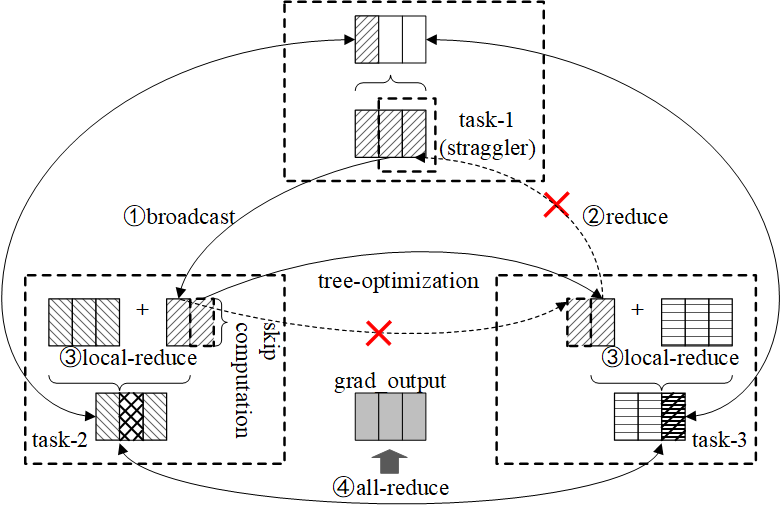}
    \caption{Illustration of the sending-collecting migration}
    \label{figure:3}
\end{figure}

Our combination selection is \emph{broadcast-reduce}, instead of \emph{scatter-gather}. The main idea behind this design is to amortize migration costs by normal tasks, and hence alleviate the burden of the straggler as much as possible. Specifically, compared with the peer-to-peer \emph{scatter}, the straggler sends the same data volume using \emph{broadcast}. The latter has a built-in tree-based optimization, which enables other normal tasks already received data as new senders to continuously transfer data. That addresses the single-point communication bottleneck problem for the already slow straggler, since connection management consumes many resources, especially when a lot of tasks as receivers are normal. For \emph{reduce} and \emph{gather}, we have the same comparison analysis where the former can further amortize the abstract summation costs. Fig.~\ref{figure:3} uses three tasks as an example to illustrate the two procedures \textcircled{\footnotesize 1} and \textcircled{\footnotesize 2}. Here task-1 as stragglers broadcasts two columns to other normal tasks but each of the latter just computes one column and skips computations related to other columns. task-2 and task-3 respectively play the roles of new sender and collector for better scalability.

Note that \emph{broadcast-reduce} enforces normal tasks to maintain redundant columns for correctness. Such additional costs can be alleviated because some data have already existed across tasks under tensor parallelism. For column-wise tensor parallelism, as shown in Fig.~\ref{figure:2}, during forward- and backward-propagations, we only need to broadcast two matrices \emph{weight} and \emph{grap\_output}. The \emph{input} matrix has already been available on everywhere. Differently, for row-wise tensor parallelism, \emph{grap\_output} is replicated among tasks but \emph{input} and \emph{weight} are split. We thereby broadcast the latter two if necessary.

Last but not least, we merge \emph{reduce} with the normally executed \emph{all-reduce} to further optimize migration costs. Recall that column-wise tensor parallelism requires an \emph{all-reduce} operation on \emph{grad\_output} at the end of backward propagation. Under data migration, as shown in Fig.~\ref{figure:3}, a task will first perform its own matrix multiplications and then process the submatrix received from stragglers. The \emph{all-reduce} operation cannot be run until all results of migrated submatrices have been collected. Clearly, these results are transferred two times respectively in \textcircled{\footnotesize 2}\emph{reduce} and \textcircled{\footnotesize 4}\emph{all-reduce}. To eliminate the redundant communication, we delay \textcircled{\footnotesize 2}\emph{reduce} and transform it from a global behavior into a local operation. That is, once a task completes its own computations and the newly-assigned workloads, it directly accumulates results of the migrated submtrix into its own matrix to perform a local reduce operation. For example, task-2 can add results into those in the second column in its own output matrix. For tasks, the location is the third column. By contrast, the straggler does nothing since the two columns have been migrated. Afterwards, we just need to run \textcircled{\footnotesize 4}\emph{all-reduce}, and \textcircled{\footnotesize 2}\emph{reduce} can be safely removed.

In row-wise tensor parallelism, we perform similar matrix computation and migration process. However, due to the fact that \emph{all-reduce} is performed in the forward propagation, merging \emph{reduce} occurs after the local \emph{output} matrix on each task is ready.

\subsection{Adaptive Allocation between Resizing and Migration}\label{section:4semi:hybrid}
Although lightweight workload migration has attempted to improve efficiency, it is still not free in runtime latency and redundant data storage. By contrast, the resizing technique has prominent time and space complexities but the accuracy seriously degrades when a large pruning ratio is used in heavily heterogeneous environments. On the other hand, the real-world heterogeneous environment is complex because both the number of stragglers and the straggling skewness can be very different and dynamically change. Thus, a smart choice is to build a hybrid solution on top of resizing and migration techniques. However, the allocation ratio between resizing and migration is a critical issue to gain optimal overall performance between efficiency and accuracy.

In fact, our migration technique is more sensitive to the number of stragglers, because each straggler replicates a full copy of its totally migrated data onto every normal task. A large number will proportionally increase the volume of redundant data, which consumes a lot of storage and bandwidth resources for normal tasks. It also weakens the benefit of tree-based communication optimizations since the number of normal tasks decreases. On the other hand, the accuracy loss of ZERO-resizing heavily depends on how much slower the straggler is. A large straggling skewness requires a big pruning ratio $\gamma$, and hence yields huge accuracy loss. We thereby discuss how to cope with the two scenarios in our hybrid solution, and particularly study the allocation ratio $\beta$.

\stitle{A Single Straggler with Heavy Straggling Skewness.}
Based on Eq.~(\ref{equation:1}), we know if ($L\!\cdot\!\gamma$) workloads/dimensions are removed, then the straggler can catch up with normal tasks. Conventionally, we can assign them among all tasks including the straggler, but the overall runtime latency can be large. Our improvement is to use resizing to partially alleviate the latency. Now the problem is how to assign these workloads between resizing and migration. Since resizing can be run only on the straggler, the principle is to balance additional costs caused by resizing and migration, respectively on the straggler and every other normal task. The former consists of two parts. One is the static space allocation overhead $\Omega_1$ for the submatrix with reduced dimensions. The other is the proportionally increased dimension extracting cost function $\Omega_2()$ for $L\!\cdot\!\gamma\!\cdot\!(1\!-\!\beta)$ data. The latter includes the communication cost function $\Phi_1()$ and computation cost function $\Phi_2()$, both of which are related to $L\!\cdot\!\gamma\!\cdot\!\beta$. As shown in Eq.~(\ref{equation:2}), we can compute $\beta$ if statistics and functions are available. $\Omega_1$ can be tested as prior-knowledge before training. For other functions, we extract several sampling points from history statistics to simulate the curve trend.

\begin{equation}
\begin{aligned}
\Omega_1 + \Omega_2\big(L\!\cdot\!\gamma\!\cdot\!(1\!-\!\beta)\big)=\Phi_1\big(L\!\cdot\!\gamma\!\cdot\!\beta\big) + \Phi_2\big(\frac{L\!\cdot\!\gamma\!\cdot\!\beta}{e-1}\big)
\end{aligned}
\label{equation:2}
\end{equation}

Because a normal task will receive a full copy of migrated data but only compute some of them, another problem is how to quickly identify the latter. Our solution is to virtually renumber normal tasks. Assume that the straggler has the original rank number $r_{k}$. For a normal task $r_{i}$, its new number is $r_{i}^{'}\!=\!(r_i\!+\!e\!-\!r_{k})\%e$. Let $L_{mig}$ be the number of migrated columns. Then each normal task will process $m\!=\!L_{mig}/(e\!-\!1)$ columns and the specific range for $r_{i}$ is [$m\!\cdot\!(r_{i}^{'}\!-\!1)$, $m\!\cdot\!r_{i}^{'}\!-\!1$]. For example, in Fig.~\ref{figure:3}, the new rank of task-2 is (2+3-1)\%3=1, and $m\!=\!1$. Then task-2 selects the first column for computation. Similarly, task-3 selects the second column.

\stitle{Multiple Stragglers with Different Straggling Skewness.}
When there are a considerable number of stragglers, the rapidly growing temporary storage overhead and redundant communication costs make it hard to apply the migration scheme to all stragglers. In this case, we categorize all stragglers into two groups. One with slight heterogeneity runs matrix resizing, and the other with heavy heterogeneity runs workload migration. We next focus on the grouping criterion.

Before detailed analysis, we first need to identify who is the straggler. In Eq.~(\ref{equation:1}) we have discussed a method of comparing the runtime $T_i$ of task-$i$ with the average $T_{avg}$. In fact, $T_{avg}$ is a compromise criterion because a strict criterion enforces many tasks to be stragglers and then a lot of dimensions are pruned. The resulting accuracy is poor. Now our hybrid solution integrates the accuracy-loss-free migration. Thus, we can use the most strict criterion, i.e., the minimum runtime $T_{min}$ among all tasks. task-$i$ with $T_i^j\!>\!T_{min}^j$ at iteration-$j$ becomes a straggler.

To determine a reasonable grouping criterion, we sort all tasks in descending order of $T_i^j$, and then detect whether or not the $x$-th task with rank $i$ should migrate partial workloads, so as to decrease its $T_i^j(x)$ to the ideal $T_{min}^j$. Such attempts are performed from the first/slowest task in a brute-force manner. In particular, given task-$i$ with the $x$-th highest runtime, we say the migration is cost-effective, if the runtime reduction ($T_i^j(x)\!-\!T_{min}^j$) is more than the maximum additional runtime cost caused by processing its migrated workloads, among other ($e\!-\!x$) tasks. Clearly, with $x$ increasing, the former decreases while the latter increases. Let $L_i^j$ indicate the current workloads of task-$i$. Assume that $x$ is the value right before cost-effectiveness. Then the total volume of migrated workloads is $\Gamma(x)=\sum_{k=1}^{x}L_i^j(x)\cdot\frac{T_k^j-T_{min}^j}{T_k^j}$. We now analyze the additional cost caused by them. The first is the total communication cost of sending migrated data and collecting results for x tasks, i.e. $\Phi_1()$ used in Eq.~(\ref{equation:2}), which is related to $\Gamma(x)$. Another is the increased computation cost. For a receiving task-$y$, it receives $\frac{1}{e-x}\!\cdot\!\Gamma(x)$ workloads. We can easily infer the computation cost since its speed $\frac{L_y^j}{T_y^j}$ is available. Eq.~(\ref{equation:3}) mathematically shows the $f$ function, and $x$ is the largest value that guarantees $f(x)\!>\!0$. That means the total $x$ tasks should run migration to reduce their runtime costs.

\begin{equation}
\begin{aligned}
f(x) &= \big(T_i^j(x)\!-\!T_{min}^{j}\big) \\
&- \Phi_1\big(\Gamma(x)\big) - \max\limits_{y\in[x+1,e]}\bigg(\frac{\Gamma(x)}{e-x}\cdot\frac{T_y^j}{L_y^j}\bigg)
\end{aligned}
\label{equation:3}
\end{equation}

Assume that there are totally $z$ stragglers. Once the migration bound $x$ is determined, we know the other $(z\!-\!x)$ tasks should run resizing to prune dimensions. Otherwise, they will slow down the overall performance, since these heavy stragglers have already attempted to reduce their runtimes to $T_{min}$ via migration. The specific pruning ratio can also be computed by Eq.~(\ref{equation:1}) but the straggling criterion $T_{avg}$ should be replaced by $T_{min}$.

\subsection{Execution of the Hybrid SEMI-migration}
Now, we present the execution of the hybrid SEMI-migration in Algorithm~\ref{algorithm:semi}. Here, $t$ represents the training at the $t$-th epoch, and $i$ is the task's ID. Firstly, we conduct a pre-test that includes two tasks (line 1). On one hand, we test the additional cost of different pruning ratios, constructing functions for $\Omega_1$ and $\Omega_2$. On the other hand, we test the communication cost of different migration ratios, constructing a function for $\Phi_1$. Subsequently, we use an all-gather operation to build a training time list, where each element $T_i$ represents the training time of task-$i$ in the previous iteration (line 2). After constructing the list, we calculate $T_{min}$ and the number of straggling tasks $z$, removing the times of normal tasks from the list (lines 3-5). Next, we determine the parameters for the migration and resizing schemes respectively (lines 7-24). If there is only a single straggler with heavy straggling skewness, we use Eq.~(\ref{equation:2}) to calculate its migration ratio and pruning ratio, and perform migration and zero-resizing accordingly (lines 7-12). Otherwise, based on Eq.~(\ref{equation:3}), we calculate the number of tasks $x$ for migration and $z-x$ for zero-resizing, and train with the parameters processed in the corresponding proportions (lines 13-24).

\begin{algorithm}
\caption{SEMI-migration Execution(task-$i$, $t$-th epoch)}
\label{algorithm:semi}
\begin{algorithmic}[1]
\STATE pretestDiffRatio($\Omega_1$, $\Omega_2$, $\Phi_1$)
\STATE $T\_list^t$ = all-gather$(T^t)$ and descendSort
\STATE $T_{min}^{t} = T_e^{t}$
\STATE $z=\sum_{k=1}^eT_k^t>T_{min}^{t}$
\STATE $T\_list^t$ = top-$z$Select($T\_list^t$)
\STATE compute $\gamma_{i}^t$ by Eq.~(\ref{equation:1})
\IF{$z=1$}
    \IF{$T_i=T_{min}$}
        \STATE compute $\beta^{t}$ by Eq.~(\ref{equation:2})
        \STATE migration($L\!\cdot\!\beta^{t}$)
        \STATE zero-resizing($L\!\cdot\!(1-\beta^{t})$)
    \ENDIF
\ELSE
    \FOR{$k=1$ to $z$}
        \STATE compute $f(k)$ by Eq.~(\ref{equation:3})
        \IF{$f(k)<0$}
            \STATE $x=k-1$ and break
        \ENDIF
    \ENDFOR
    \IF{$T_i^t>=T_x^t$}
        \STATE migration($L\!\cdot\!\gamma_{i}^{t}$)
    \ELSE
    \STATE zero-resizing($L\!\cdot\!\gamma_{i}^{t}$)
    \ENDIF
\ENDIF
\end{algorithmic}
\end{algorithm}

\section{EXPERIMENTS}\label{section:5exp}
In this section, we evaluate the performance of our proposals. We implement them on the newly released platform Colossal-AI~\cite{r5}. It integrates tensor parallelism and other parallel optimizations for efficiently training foundation models.

\subsection{Experimental Setups}\label{section:5exp:setting}
We first introduce the general experimental settings for better understanding our evaluation results.

\stitle{Compared Solutions.}
We report the performance of our resizing technique (Sec.~\ref{section:3zero}, termed as ZERO) and the hybrid solution (Sec.~\ref{section:4semi}, termed as SEMI). More specifically, we test three variants of the former, including the basic ZERO-\emph{Rd} where pruned data are randomly selected, the improved ZERO-\emph{Pri} with prioritized selection based on importance evaluation, and the final ZERO-\emph{PriDiff} further enhanced by customizing the volume of pruned data across different layers. The competitor here is the 1D tensor parallelism provided by Colossal-AI\footnote{Colossal-AI also provides other multi-dimension tensor parallelism implementations for communication reduction, but they only work on the special number of GPUs. We thereby test the general-purpose 1D tensor parallelism. Nevertheless, our techniques are also suitable to other variants.}, denoted by \emph{Baseline}.

\stitle{Hardwares and Simulation Testbed.}
All tests are conducted on a server running CentOS Linux 7 (Core). It is equipped with Intel(R) Xeon(R) Gold 6126 CPU @2.60GHz, 180GB RAM; and 8 NVIDIA Tesla V100 GPUs, each of which has 16GB HBM2 and offers 112 teraFLOPS compute power by 640 Tensor Cores. All GPUs are connected by PCIe 3.0 and work with CUDA Toolkit 11.7. That naturally forms homogeneous environments. However, our proposals mainly target heterogeneous environments. On the other hand, existing studies have concluded that it is hard to accurately distinguish massive and dependent straggling factors, even though they indeed exist in practice~\cite{r15,r17,fsp}. We thereby simulate the heterogeneous testbed by injecting sleeping operations to suspend threads, for GPUs manually selected as stragglers. We particularly quantify the {\bf straggling skewness} by $\chi$, namely the simulated matrix multiplication in linear projections and transformations is $\chi$ times slower than the normal version.

\stitle{Benchmark Models and Datasets.}
Our benchmark foundation model is the built-in ViT~\cite{vit} example provided by Colossal-AI, which has recently emerged as a competitive alternative to existing representative CNN in image understanding applications. By customizing hyper-parameters (the layer number and the hidden size), we have two variants that can be fit in our total GPU memory. One is ViT-1B with 1.2 billion of parameters, and the other is ViT-3B with 2.7 billion of parameters. The dataset is CIFAR-10 consisting of 60,000 32$\times$32 color images labeled with one of 10 exclusive classes\footnote{https://www.cs.toronto.edu/~kriz/cifar.html}.

\stitle{Evaluation Metrics.}
Because our main idea is to trade accuracy for efficiency, we then focus on two important metrics, the training runtime \emph{RT} and the model accuracy \emph{ACC}. It is worth noting that the absolute accuracy depends on many fine-tuned hyper-parameters, like learning rate, epoch number, batch size, and even the volume of datasets. These complex configurations are beyond the scope of this paper, since we primarily concern the range of accuracy variation under different proposals, instead of the absolute value. Here, these configurations just need to guarantee that we can collect metric values in reasonable runtime costs. Guided by this purpose, even though the foundation models have billions of parameters, we still train them using a moderately-sized dataset. For each test, we totally run 150 epochs with batch sizes 64 and 32 respectively for ViT-1B and ViT-3B, and the uniform learning rate 0.003. We define \emph{RT} as the averaged elapsed time of an epoch, since training workloads do not change across epochs. We keep track of \emph{ACC} by separately evaluating the model after each epoch.

\subsection{Evaluation on ZERO}\label{section:5exp:zero}
Because dimension reduction can generally decrease the computation workloads, we explore its impact in both homogeneous and heterogeneous environments.

\begin{figure}[htbp]
	\centering
    \renewcommand{\thesubfigure}{\scriptsize (\alph{subfigure})\space}
	\subfigure[Model accuracy]{
			 \includegraphics[width=0.48\linewidth]{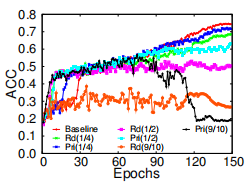}
	}%
	\subfigure[Training efficiency]{
			 \includegraphics[width=0.48\linewidth]{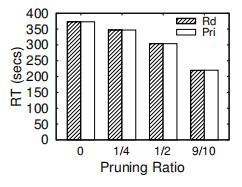}
	}
	\centering
	\caption{Overall performance in homogeneous environments (ViT-1B)}
	\vspace{-0.2cm}
	\label{figure:homo1b}
\end{figure}

\begin{figure}[htbp]
	\centering
    \renewcommand{\thesubfigure}{\scriptsize (\alph{subfigure})\space}
	\subfigure[Model accuracy]{
			 \includegraphics[width=0.48\linewidth]{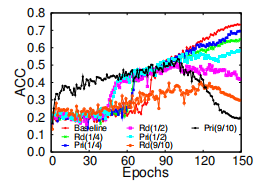}
	}%
	\subfigure[Training efficiency]{
			 \includegraphics[width=0.48\linewidth]{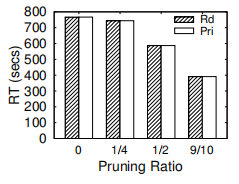}
	}
	\centering
	\caption{Overall performance in homogeneous environments (ViT-3B)}
	\vspace{-0.2cm}
	\label{figure:homo3b}
\end{figure}

\stitle{Homogeneous Evaluation.} We evaluate the overall Performance scalability in both metrics, by varying the ratio of pruned dimensions to original ones on every GPU. We test three representative {\bf pruning ratios}, i.e., $\gamma\!=\!\frac{1}{4}, \frac{1}{2}$, and $\frac{9}{10}$. Fig.~\ref{figure:homo1b} and Fig.~\ref{figure:homo3b} compare \emph{Baseline}, ZERO-\emph{Rd}, and ZERO-\emph{Pri}, respectively on ViT-1B and ViT-3B. \emph{PriDiff} is excluded because it only makes sense in heterogeneous environments. With the ratio being increased, in all cases we can observe that \emph{RT} gradually decreases but at expense of the growing accuracy loss. Compared with ZERO-\emph{Rd}, ZERO-\emph{Pri} narrows down the loss by up to 18\% ($\gamma\!=\!\frac{9}{10}$, Vit-1B), with nearly-zero runtime penalty. That stems from efficiently and effectively selecting unimportant dimensions for pruning. Taking $\gamma\!=\!\frac{1}{2}$ as an example, ZERO-\emph{Pri} generates up to 23.5\% runtime improvement with 14.9\% accuracy loss, on ViT-3B.

\begin{figure}[htbp]
	\centering
    \renewcommand{\thesubfigure}{\scriptsize (\alph{subfigure})\space}
	\subfigure{
			 \includegraphics[width=0.5\linewidth]{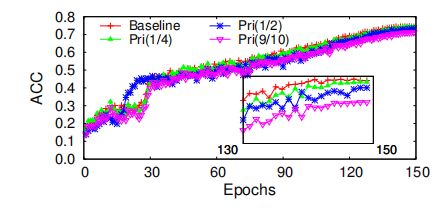}
	}
	\centering
	\caption{Accuracy variation with fixed straggling skewness (ViT-1B)}
	\vspace{-0.2cm}
	\label{figure:hetefixed1b}
\end{figure}

\begin{figure}[htbp]
	\centering
    \renewcommand{\thesubfigure}{\scriptsize (\alph{subfigure})\space}
	\subfigure{
			 \includegraphics[width=0.5\linewidth]{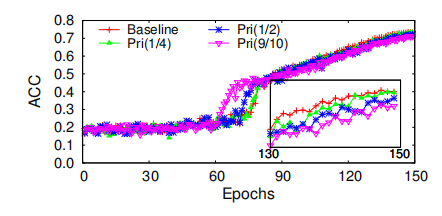}
	}
	\centering
	\caption{Accuracy variation with fixed straggling skewness (ViT-3B)}
	\vspace{-0.2cm}
	\label{figure:hetefixed3b}
\end{figure}

\stitle{Heterogeneous Evaluation.} In this suite of experiments, we simulate the dynamic heterogeneous scenario by injecting sleeping operations into different GPUs among epochs, in a round-robin manner. Thus, at any time, only one GPU acts as the straggler. Since \emph{Pri} is always better than \emph{Rd} as validated in homogeneous environments, now we remove \emph{Rd} for brevity. Below, we first demonstrate how ZERO-\emph{Pri} scales when the pruning ratio $\gamma$ varies, with straggling skewness fixed as $\chi\!=\!2$. Together with Figs.~\ref{figure:homo1b}(a) and ~\ref{figure:homo3b}(a), in Figs.~\ref{figure:hetefixed1b} and \ref{figure:hetefixed3b}, we clearly see that the accuracy loss is reduced, since now ZERO-\emph{Pri} prunes data only on the selected straggler, instead of the total 8 GPUs in homogeneous tests.

\begin{figure}[htbp]
	\centering
    \renewcommand{\thesubfigure}{\scriptsize (\alph{subfigure})\space}
	\subfigure[Model accuracy]{
			 \includegraphics[width=0.48\linewidth]{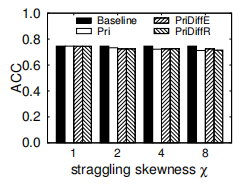}
	}%
	\subfigure[Training efficiency]{
			 \includegraphics[width=0.48\linewidth]{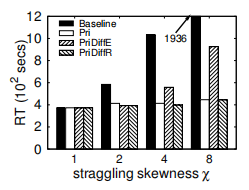}
	}
	\centering
	\caption{Overall performance in heterogeneous environments (ViT-1B)}
	\vspace{-0.2cm}
	\label{figure:hetedyn}
\end{figure}

Fig.~\ref{figure:hetedyn} further depicts how \emph{Pri} and \emph{PriDiff} scale when varying $\chi$ from 0 (homogeneous) to 8. In particular, we have two branches of \emph{PriDiff} based on how to determine the uniform pruning ratio $\gamma_i^j$. We can set it as an empirical value $\frac{1}{2}$ according to results in Figs.~\ref{figure:hetefixed1b} and \ref{figure:hetefixed3b}; or the value given in Eq.~(\ref{equation:1}). They are separated by suffixes ``E'' and ``R''. As expected, \emph{RT} of \emph{Baseline} linearly increases with $\chi$. By contrast, ZERO-\emph{Pri} guarantees that the straggler can roughly catch up with other normal tasks, and hence the overall \emph{RT} kepdf steady. In the case of $\chi\!=\!8$, the speedup of \emph{Pri} compared against \emph{Baseline} is up to 3.5, with acceptable 1.3\% accuracy loss. \emph{PriDiffE} and \emph{PriDiffR} have different favorites. The former can trade efficiency for accuracy loss reduction. When $\chi\!=\!8$, the loss is reduced to 0.3\% but the speedup factor also decreases to 2.1. Differently, compared with \emph{Pri}, the latter as a preferred enhancement can further reduce \emph{RT} by up to 4\% ($\chi\!=\!2$), and still offers comparable and even better accuracy.

\subsection{Evaluation on SEMI}\label{section:5exp:semi}
Now we design experiments to verify the effectiveness of our hybrid solution. But before that, we first investigate the effectiveness of lightweight migration optimizations. Table~\ref{tab:migration} lists the comparison of runtime per epoch, when \emph{broadcast-reduce} and \emph{scatter-gather} are respectively used to migrate partial dimensions from selected GPUs to others. For in-depth analysis, the volume of migrated data measured by $\gamma$ (see Sec.~\ref{section:5exp:zero}) ranges from 0 to 1. Besides, the number of selected GPUs $\nu$ also varies from 1 to 4. All tests are run in homogeneous environments. As reported, \emph{broadcast-reduce} outperforms \emph{scatter-gather}, because its built-in tree-based broadcasting/reducing manner yields multiple de facto senders/receivers (instead of one in \emph{scatter}). Our \emph{reduce}-merging design also contributes to its prominent efficiency. However, with the increase of $\nu$, the number of target GPUs decreases. The additional communication costs in \emph{scatter} are less significant. Thus, their performance gap narrows down.

\begin{table}[htbp]
\centering
\caption{Runtime comparison of different migration policies (secs)}
\label{tab:migration}
{\small
\begin{tabular}{|l|r|r|r|r|r|} \hline
{\bf Policies}($\boldsymbol{\nu}$)~/~$\boldsymbol{\gamma}$& {\bf 0.00}& {\bf 0.25}& {\bf 0.50}& {\bf 0.75}& {\bf 1.00}\\ \hline\hline
\emph{broadcast-reduce}(1)& 373& 425& 451& 473& 500\\ \hline
\emph{scatter-gather}(1)& 373& 478& 627& 796& 963\\ \hline
\emph{broadcast-reduce}(4)& 373& 737& 846& 998& 1,113\\ \hline
\emph{scatter-gather}(4)& 373& 779& 971& 1,198& 1,436\\ \hline
\end{tabular}
}
\end{table}

To demonstrate how the cost-benefit model (Sec.~\ref{section:4semi:hybrid}) works, we next give detailed behavior analysis in complex scenarios. Here \emph{MIG} stands for the balancing solution solely depending on data migration (Sec.~\ref{section:4semi:migration}). Also, to clearly show the difference about accuracy, we do not show the absolute accuracy value. Instead, we use \emph{Baseline} as the standard marked with zero, and then report the accuracy variation of every other solution.

\begin{figure}[htbp]
	\centering
    \renewcommand{\thesubfigure}{\scriptsize (\alph{subfigure})\space}
	\subfigure[Model accuracy]{
			 \includegraphics[width=0.48\linewidth]{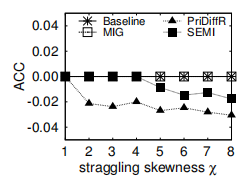}
	}%
	\subfigure[Training efficiency]{
			 \includegraphics[width=0.48\linewidth]{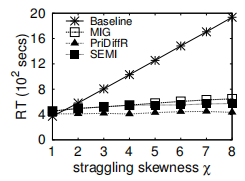}
	}
	\centering
	\caption{Scalability with a single straggler (ViT-1B)}
	\vspace{-0.2cm}
	\label{figure:scaone}
\end{figure}

Fig.~\ref{figure:scaone} plots \emph{ACC} and \emph{RT} against the straggling skewness $\chi$, if only one straggler happens. As shown in subfigure(b), \emph{RT} of \emph{Baseline} linearly increases with $\chi$ because of waiting costs. \emph{MIG} mitigates the efficiency degradation but is still far from ideal, due to the migration cost, especially when $\chi$ is large. By contrast, ZERO-\emph{PriDiffR} and SEMI both exhibit prominent scalability. However, subfigure(a) reveals that the former causes much more accuracy loss than the latter. In general, SEMI achieves an overall success in terms of efficiency and accuracy, by smartly combining ZERO and \emph{MIG} techniques.

\begin{figure}[htbp]
	\centering
    \renewcommand{\thesubfigure}{\scriptsize (\alph{subfigure})\space}
	\subfigure[Model accuracy]{
			 \includegraphics[width=0.48\linewidth]{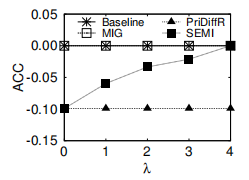}
	}%
	\subfigure[Training efficiency]{
			 \includegraphics[width=0.48\linewidth]{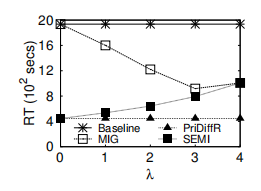}
	}

	\centering
	\caption{Scalability with multi-stragglers (ViT-1B)}
	\vspace{-0.2cm}
	\label{figure:scamore}
\end{figure}

We finally evaluate SEMI when half of the total eight GPUs become stragglers but their $\chi$ settings vary from 8 down to 2 with step size 2. Recall that SEMI will sort stragglers in descending order of $\chi$ and then adaptively divide them into two parts, to respectively run \emph{MIG} and ZERO-\emph{PriDiffR}. To find a real optimal division, we repeatedly run SEMI by manually varying the number of GPUs running \emph{MIG}, i.e. $\lambda$. Fig.~\ref{figure:scamore} reports the two metrics versus $\lambda$ plots. We clearly see that at the extreme cases of $\lambda\!=\!0$ and 4, SEMI respectively degrades to ZERO-\emph{PriDiffR} and \emph{MIG}. Our analysis can help it to roughly hit the ``sweet spot'' ($\lambda\!=\!3$) where we can efficiently train the model with small accuracy penalty.

\section{Related Works}\label{section:6related}
This section first overviews parallel deep training solutions, followed by summarizing efforts in heterogeneous environments, and then outlines  works of model compression.

\subsection{Parallel Deep Learning}\label{section:6related:systems}
Data parallelism and model parallelism are two mainstream parallel training solutions. The former distributes sample data across devices to increase the throughput~\cite{dataparallel}, but parameters must be fully replicated everywhere, which can easily exhaust memory resources especially for foundation models with billions of parameters. This drawback can be overcome by the latter which divides parameters, rather than samples. It has two specific branches, pipeline parallelism performs layer-centric parameter divison where intermediate data between layers are transferred across devices~\cite{r3}, yielding underutilization of compute power due to the bubble overhead; tensor parallelism splits tensors along specified dimensions and then trains models via distributed matrix multiplication~\cite{r4}, that makes full use of devices. Some recent works propose to partition samples and tensors together for better scalability~\cite{r1,r2}. The recently released open-source hybrid platform  Colossal-AI~\cite{r5} also integrates all solutions mentioned above. Among them, tensor parallelism always plays a key role. However, its built-in all-reduce requirements yield frequent synchronization operations, and hence slow down convergence speed, especially in heterogeneous environments.

\subsection{Efforts in Heterogeneous Environments}\label{section:6related:heterogeneous}
Heterogeneous environments are usually encountered in practice. There is a flurry of efforts for the resulting straggling problem, but most of them focus on data parallelism. Some researchers attempt to relax consistency constraints. Towards this end, they confine the synchronization barrier to a subset of tasks running fast and others as stragglers are directly skipped~\cite{r6,r7,r8}, which wastes compute power. Another alternative is stale synchronous parallel~\cite{r11,r12,r13}, or even asynchronous parallel without any barrier~\cite{r9,r10}. In these works fast tasks continuously proceed to a few, or infinite number of subsequent synchronizing points. That mitigates the blocking waiting problem but generates different parameter versions among tasks. However, for tensor parallelism, it will lead significant errors accumulated by frequent inter-layer synchronization operations. Besides, dynamic data migration as a classic solution can solve the straggling problem without any accuracy loss~\cite{r14,r15,r16}. But the key issue is to reduce migration costs. Existing optimizations include pre-replicating samples~\cite{r17} and overlapping normal computations with data migration, since only a few samples are used in a batch~\cite{fsp}. We also follow the migration idea but abandon corresponding optimizations, because the former challenges the limited GPU memory capacity; while the latter does not work for tensor parallelism where tensors are fully involved in computations. Instead, we give our own lightweight optimization by carefully selecting communication primitives and merging some of them. Note that the partial-use feature of samples also motivates researchers to tune workloads on demand by customizing batch size settings for different tasks~\cite{r19,r20,r21,fsp}. Under this design, samples on fast tasks will be scanned more frequently, that yields skewed contributions to parameter refinement, and even makes the model overfit to these samples. Similarly, this cannot be extended to tensors. Even for samples, it also cannot work because customizing batch sizes will change the size of output tensors and hence lead to the all-reduce communication failure.

With the rapid development of foundation models, recently workload balancing for model parallelism in heterogeneous environments has gained increasing attention. Park et al. logically group tasks where pipeline parallelism is used with a group and data parallelism is used across groups~\cite{r22}. At the very beginning of training, they differentiate the group size to react to heterogeneity. Other works focus on the skewed initial partitioning by considering different network bandwidths and compute powers~\cite{r24,r25,r26}. However, all of these techniques mentioned above are static, that cannot cope with dynamic stragglers. We are also aware that overlapping computations and communications can improve training efficiency and partially eliminate the negative impact of heterogeneous compute power~\cite{r31}. Colossal-AI also integrates this design. However, the effectiveness in practice is marginal based on our tests, since such overlapping/asynchronizing optimization only works within a layer for correctness.

\subsection{Compressing Model for Workload Reduction}\label{section:6related:others}
The core idea of our proposals is to dynamically control workloads by resizing matrices. We know that there also exists another path to workload reduction, i.e., model compression. That is extensively studied in marginal inference, because the reduced model can be fit into low-configured devices. Towards this end, researchers prune models by removing unimportant structures~\cite{r27,r28} or masking parameters with small absolute values~\cite{r30}. However, these aggressive pruning operations are irrevocable, which cannot react to dynamic environments. Worse, the pruned data must be determined by in-depth analysis on well-trained parameters and weights. But during training, such prior-knowledge is not available.

\section{Conclusion}\label{section:7con}
This paper studies the straggling issue of tensor parallelism, which recently emerges when training foundation models in complex heterogeneous environments. We propose to resize matrices involved in core tensor computations with priority enhancement, and data migration with lightweight optimizations. Both can dynamically balance workloads to reduce waiting costs, but they have different advantages in terms of efficiency and accuracy. We thereby design a hybrid solution by combining them, which achieves an overall success, as validated in experiments.

\bibliographystyle{IEEEtran}
\bibliography{reference}

\end{document}